\documentclass{article}
\usepackage{emulateapj}
\usepackage{apjfonts}
\usepackage{graphics}

\newenvironment{inlinefigure}{%
\def\@captype{figure}%
\noindent\begin{minipage}{0.999\linewidth}\begin{center}}
{\end{center}\end{minipage}\smallskip}

\newlength{\colwidth}
\newcommand{\Mpc}{\hbox{{\rm Mpc}}}
\newcommand{\kpc}{\hbox{{\rm kpc}}}

\newcommand{\hydra}{{\sc{HYDRA}}}

\newcommand{\vpfit}{{\sc{VPFIT}}}
\newcommand{\cloudy}{{\sc{CLOUDY}}}

\newcommand{\lya}{{\mbox Ly$\alpha$}}
\newcommand{\Cfour}{{\rm C\,{\sc IV}}}
\newcommand{\Nfive}{{\rm N\,{\sc V}}}

\newcommand{\Osix}{{\rm O\,{\sc VI}}}
\renewcommand{\H}{\ion{H}{1}}

\newcommand{\He}{\ion{He}{1}}
\newcommand{\Hep}{\ion{He}{2}}

\newcommand{\h}{{\rm H\,{\sc I}}}

\newcommand{\hep}{{\rm He\,{\sc II}}}

\newcommand{\cfour}{{\rm C\,{\sc IV}}}

\newcommand{\be}{\begin{equation}}
\newcommand{\ee}{\end{equation}}
\newcommand{\cm}{\hbox{\rm cm}}

\newcommand{\K}{\hbox{\rm K}}
\newcommand{\kms}{\hbox{{\rm km}\,{\rm s}$^{-1}$}}
\newcommand{\ltsima}{\mbox{$\; \buildrel < \over \sim \;$}}
\def \simlt{\lower.5ex\hbox{\ltsima}}            
\def \gtsima{\mbox{$\; \buildrel > \over \sim \;$}}
\def \simgt{\lower.5ex\hbox{\gtsima}}            
\lefthead{Theuns et al.}
\righthead{Galactic winds in the IGM}

\begin{document}


\title{Galactic winds in the Intergalactic Medium\altaffilmark{1}}
\author{Tom Theuns\altaffilmark{2,3}, Matteo Viel\altaffilmark{4,2},
Scott Kay\altaffilmark{5}, Joop Schaye\altaffilmark{6}, Robert F.
Carswell\altaffilmark{2}, and Panayiotis Tzanavaris\altaffilmark{2}}

\altaffiltext{1}{Based on observations made at the W.M. Keck Observatory
which is operated as a scientific partnership between the California
Institute of Technology and the University of California; it was made
possible by the generous support of the W.M. Keck Foundation. }
\altaffiltext{2} {Institute of Astronomy, Madingley Road, Cambridge CB3
0HA, UK}
\altaffiltext{3} {Universitaire Instelling Antwerpen, Universiteitsplein
1, B-2610 Antwerpen, Belgium}
\altaffiltext{4} {Dipartimento di Fisica \lq Galileo Galilei\rq\, via
Marzolo 8, I-35131 Padova, Italy}
\altaffiltext{5} {Astronomy Centre, CPES, University of Sussex, Falmer, 
Brighton BN1 9QJ, UK}
\altaffiltext{6} {School of Natural Sciences, Institute for Advanced Study, Einstein Drive, Princeton NJ 08540}

\begin{abstract}
We have performed hydrodynamical simulations to investigate the effects
of galactic winds on the high-redshift ($z=3$) universe. Strong winds
suppress the formation of low-mass galaxies significantly, and the
metals carried by them produce \Cfour{} absorption lines with
properties in reasonable agreement with observations. The winds have
little effect on the statistics of the \h{}-absorption lines, because
the hot gas bubbles blown by the winds fill only a small fraction of
the volume and because they tend to escape into the voids, thereby
leaving the filaments that produce these lines intact.
\end{abstract}

\keywords {cosmology: observations --- cosmology: theory ---
galaxies: formation --- intergalactic medium --- quasars: absorption
lines}

\section{Introduction}
Feedback from star formation is thought to play an important role in
the formation of galaxies. For example, theoretical models require
feedback in order not to overproduce the number of low mass galaxies
(White \& Rees 1978) and the fraction of baryons that cools (e.g., Balogh
et al.\ 2001). Models without feedback also predict disk galaxies that
are too small (e.g., Navarro \& Steinmetz 1997) and an X-ray background
that is too strong (e.g., Pen 1999).

Observations of galactic winds indicate that strong feedback processes
do indeed occur. Observations in X-rays, and of optical and UV lines,
indicate that most local starbursts (e.g., Heckman 2000), as well as
high-redshift Lyman Break galaxies (e.g., Pettini et al.\ 2001), drive
winds with a mass loss rate comparable to their star formation
rate. The wind speeds are high, 100-1000~\kms, and so the conversion of
supernova (SN) energy into kinetic energy must be quite efficient.
This process is currently not well understood because of the
complications that arise from the multiphase nature of the interstellar
medium (ISM, e.g., McKee \& Ostriker 1977; Efstathiou 2000). Simulations
that take some of these complications into account, have shown that SNe
can indeed plausibly power a wind (e.g., Mac Low \& Ferrara 1999).

Although successful models of galaxies appear to require feedback, the
same is not true for models of the high-redshift ($z \ga 2$)
intergalactic medium (IGM). The IGM can be studied in great detail from
the properties of the hydrogen absorption lines, seen in the spectra of
background quasars (see Rauch 1998 for a review). Hydrodynamical
simulations of the IGM (see Efstathiou, Schaye \& Theuns 2000 for a
recent review) as well as semi-analytic (Bi \& Davidsen 1997; Viel et
al.\ 2002) and analytic (Schaye 2001) models, have been very successful
in reproducing the statistical properties of the observed \h{}
lines. These models, which generally do \emph{not} take feedback into
account (but see Cen \& Ostriker 1999), suggest that the Ly$\alpha$
forest absorption arises in a network of voids and filaments, with the
higher column density absorbers located at the intersections of
filaments. The low column density absorbers are extended structures
with densities around the cosmic mean, which contain a large fraction
of the baryons in the universe.

Although feedback has so far largely been ignored in models of the IGM,
the detection of metals in the IGM suggests that it {\em does} play a
role. The higher column density ($N_\h \ga 10^{14.5}~\cm^{-2}$)
Ly$\alpha$ absorption systems generally have detectable absorption by
\Cfour\ (Cowie et al.\ 1995) and, at least at $z\la 2.5$, \Osix\
(Carswell, Schaye, \& Kim 2002). Furthermore, there is statistical
evidence that metals are also present at somewhat lower densities
(Cowie \& Songaila 1998; Ellison et al.\ 2000; Schaye et al.\
2000a). Simple photo-ionization models indicate that the absorbers have
a metallicity of order 0.1 -- 1 per cent solar (e.g., Cowie et al.\
1995; Rauch, Haehnelt, \& Steinmetz 1997; Hellsten et al.\ 1997;
Carswell et al.\ 2002).

If galactic winds are ubiquitous and able to transport mass to
large distances, then they may be responsible for enriching the IGM
with metals. Indeed, numerical simulations of galactic winds in a
cosmological setting (e.g., Gnedin 1998; Cen \& Ostriker 1999; Aguirre
et al.\ 2001a, 2001b; Thacker, Scannapieco, \& Davis 2002; Springel \&
Hernquist 2002) suggest that winds could enrich a substantial volume
fraction of the IGM to the inferred levels.

Therefore, both observational and theoretical considerations suggest
that some fraction of galaxies may undergo an episode in which they
blow a strong wind into their surroundings. This may have observational
effects on the \lya{} forest in quasar spectra (Theuns, Mo, \& Schaye
2000; Croft et al.\ 2002), which may have already been detected (Rauch
et al.\ 2001; Adelberger et al.\ 2002).  In this {\em Letter}, we use
hydrodynamical simulations to investigate whether galactic winds that
strongly influence the properties of small galaxies, and are effective
in enriching the IGM with metals, can do so without undermining the
success of current models of the IGM.
\begin{figure*}
\label{fig:reion}
\centerline{\resizebox{1.92\colwidth}{!}{\includegraphics{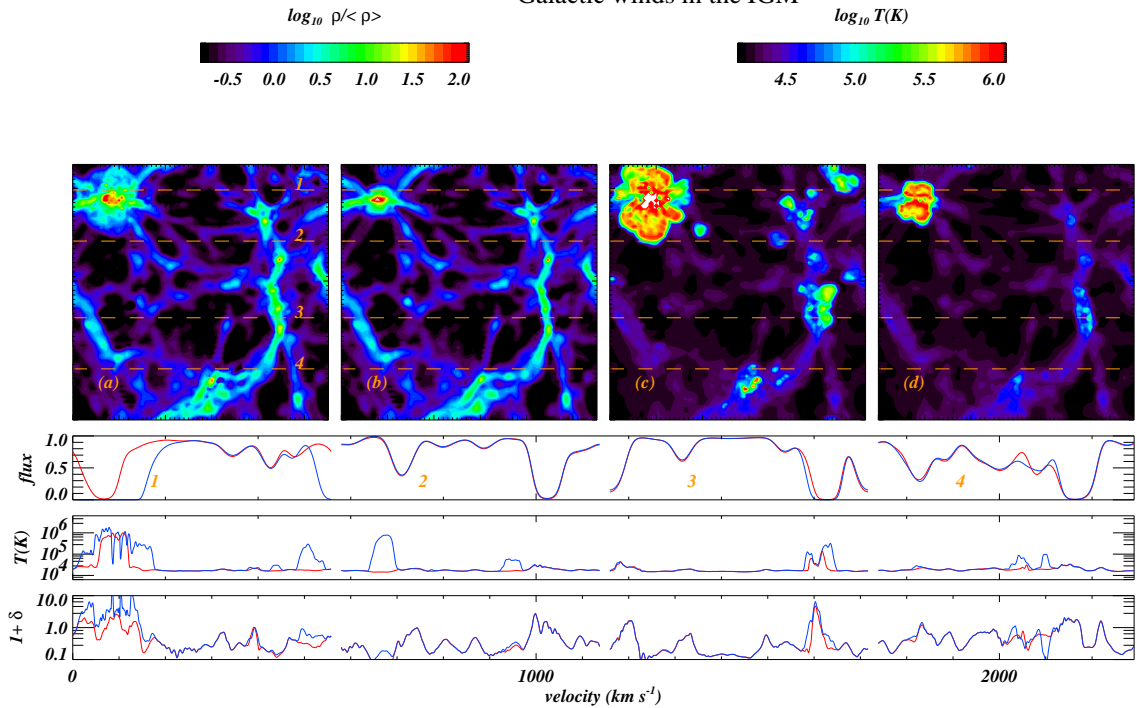}}}
\figcaption{
{\it Top panels:} Density (a, b) and temperature (c, d) for a slice
through the simulation with feedback (a, c) and the simulation without
feedback (b, d), at redshift $z=3$. The simulation box is 5.0$h^{-1}$ (co-moving) \Mpc{}. The slice
is chosen to go through the most massive object in the simulation. In
the feedback simulation, hot bubbles of gas surround the
galaxies. These tend to expand into the voids, thereby leaving the
filamentary network of higher density regions unaffected. The four
numbered horizontal lines (from top to bottom) are sight lines for
which the \lya{} spectrum, temperature, and density, are shown in the
bottom panel (from left to right, offset for clarity). The feedback and
no feedback cases are shown as blue and red lines, respectively. Only
the strongest lines are significantly affected by feedback.}
\end{figure*}

\section{Hydrodynamical simulations}
We have performed simulations of vacuum energy dominated,
cosmologically flat cold dark matter models, with cosmological
parameters $(\Omega_m,\Omega_bh^2, h,
\sigma_8,Y)$=(0.3,0.019,0.65,0.7,0.24). The lower value of $\sigma_8$
was chosen in view of the recent results from the Two-Degree Field
survey (Lahav et al.\ 2002). The simulations are performed using a
modified version of \hydra{} (Couchman, Thomas \& Pearce 1995; Theuns
et al.\ 1998), in which the gas is photo-heated by an imposed uniform
UV-background which evolves with redshift and reionizes \H\ and \He\
at $z\sim 6$ and \Hep\ at $z\sim 3.2$. The photoheating rates were
adjusted in order to match the temperature measurements of Schaye et
al.\ (2000b), where the current simulation is referred to as the \lq
designer\rq~ simulation.  Non-equilibrium rates for photoionization,
cooling, and heating are computed using the fits in Theuns et al.\
(1998). The particle masses for dark matter and baryons are $6.5$ and
$1.1\times 10^6~M_\odot$ respectively, and the simulation box (which
has periodic boundary conditions) is 5.0$h^{-1}$ (co-moving) \Mpc{} on
a side. The Plummer softening is 5$h^{-1}$ (co-moving) \kpc{}.

Star formation and feedback are implemented using the prescription
described in detail in Kay et al.{} (2002). Briefly, cold gas with
density $\rho/\langle\rho\rangle > 1000$ and temperature $T<50,000$~K
is converted into collisionless stars.  Each star particle heats the
nearest gas particle using equation (3) of Kay et al. (2002), with a
given efficiency parameter (we use $\epsilon=1$). Reheated gas is
prohibited from cooling for $10^7$ yrs to crudely account for the
multi-phase nature of the ISM. This simple feedback prescription has
the desired effect of blowing hot bubbles around star-forming
regions. The heated gas is also enriched with metals. A yield of three
times solar matches the metal line data well (see below).

\begin{inlinefigure}
\label{fig:reion}
\centerline{\resizebox{0.96\colwidth}{!}{\includegraphics{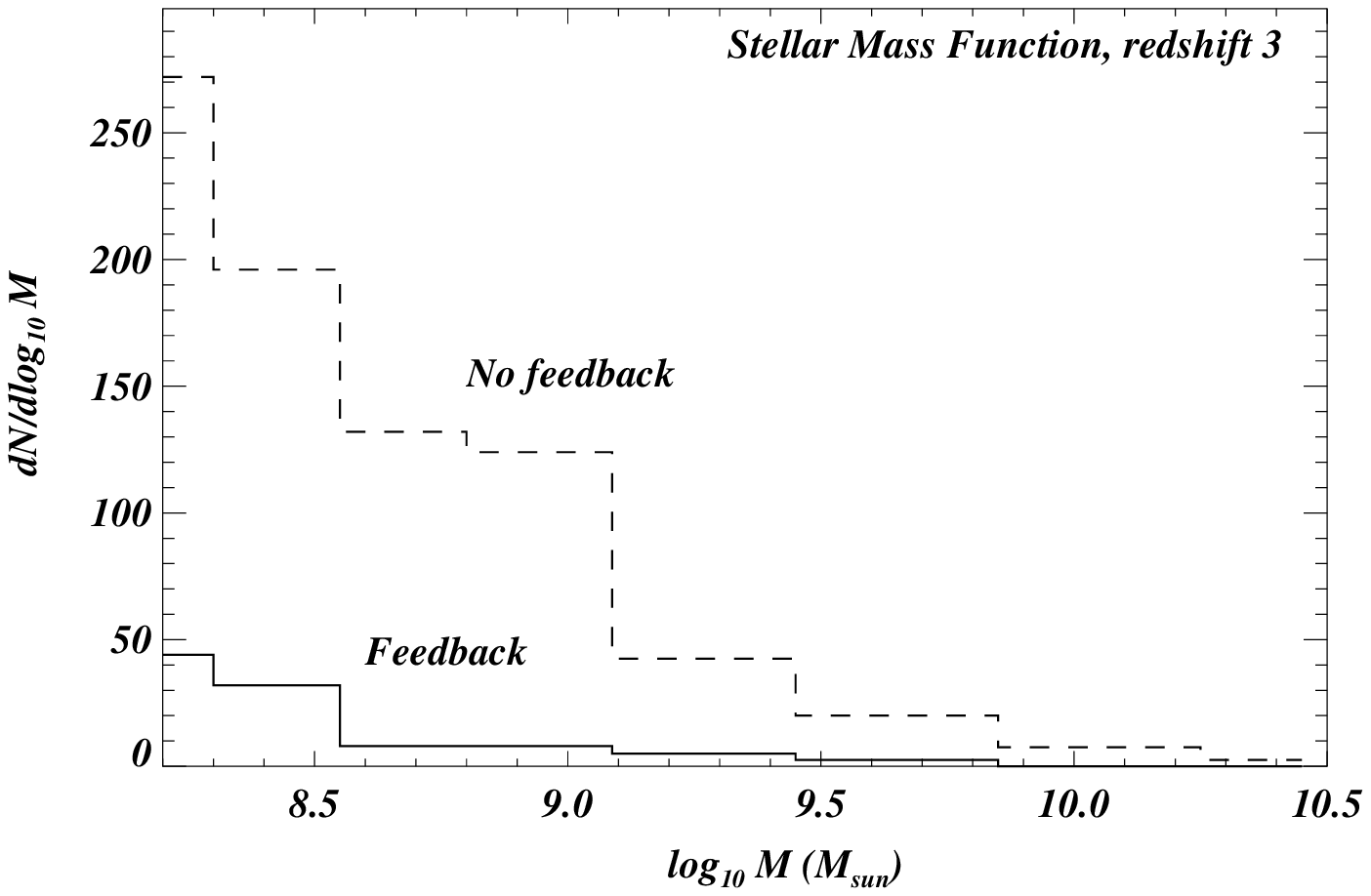}}}
\figcaption{
Stellar mass function of galaxies at redshift $z=3$, for the simulation
with feedback (full line) and without feedback (dashed line). The
feedback scheme has a dramatic effect on the masses of the galaxies.}
\end{inlinefigure}

\begin{figure*}
\label{fig:reion}
\centerline{\resizebox{1.92\colwidth}{!}{\includegraphics{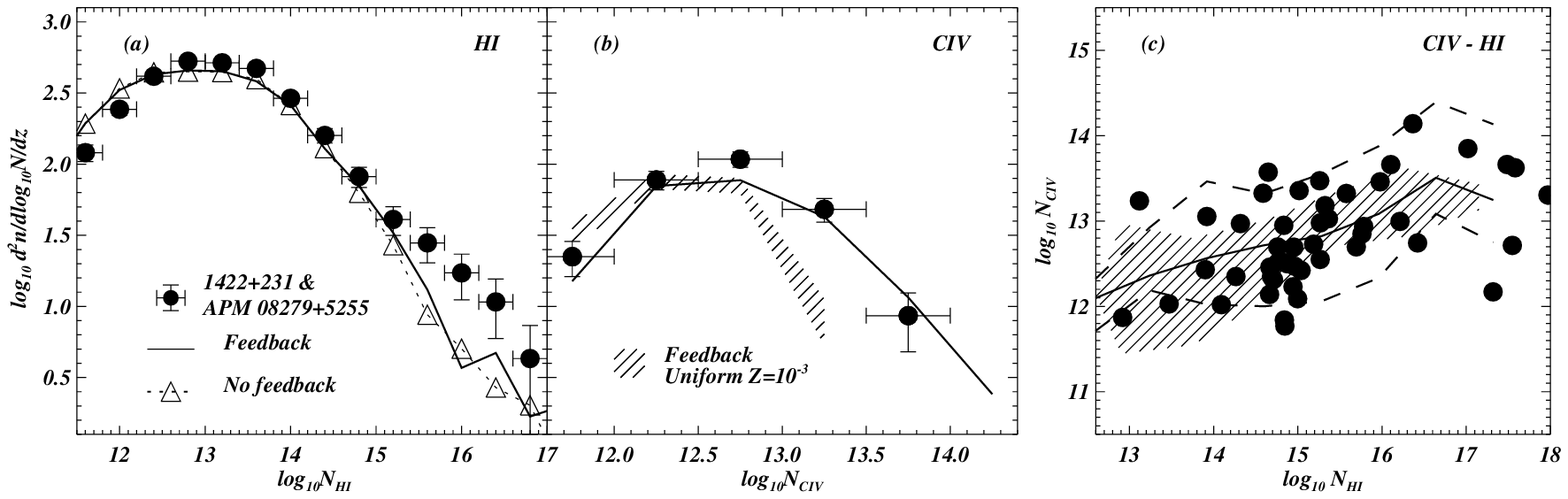}}}
\figcaption{
Column density distribution functions (CDDFs) of \H\ (panel a), \Cfour\
(panel b), and the \Cfour{} versus \H{} column density of systems
(panel c). Filled circles refer to the combined line-lists of quasars
Q1422+231 (Ellison et al.\ 2000; $z_{\rm em}=3.6$) and APM 08279+5255
($z_{\rm em}=3.91$), and the solid line indicates the results for the
feedback simulation at $z=3$. Panel (a): The \H\ CDDF of the feedback
simulation is nearly identical to that of the simulation without
feedback (dotted line and triangles), and both fit the observations
very well for $\log N_{HI} \la 15$. Panel (b): The feedback simulation
matches the observed \Cfour\ CDDF, but imposing a uniform metallicity
($Z=10^{-3}Z_\odot$, hashed region) does not work so well. Panel (c):
\Cfour{} versus \h{} column density, for systems identified on a
smoothing scale of $\pm 150\kms${}. The median for the feedback
simulation is indicated with the full line, with dashed lines
indicating the 5 and 95 per centiles. The hashed region is the
corresponding range for the uniform metallicity case. The feedback
simulation reproduces the observed median and scatter better than the
uniform $Z$ case.}
\end{figure*}

\noindent We make no attempt to model the transition from SN bubble to
galactic wind correctly, since our simulations lack both in resolution
and input physics. We do not include cooling due to metal transitions
which could be important. Our primary aim in this preliminary
investigation, is to test whether strong feedback in the form of
galactic winds can coexist with the \lya{} forest as we know it from
simulations without feedback. We use the galaxy mass function and the
properties of the \Cfour\ forest to demonstrate that the implemented
feedback is indeed significant. We have performed two simulations with
identical initial conditions, one with and one without feedback, so
that we can examine its effect directly.

\section{Results}
\begin{inlinefigure}
\label{fig:reion}
\centerline{\resizebox{0.96\colwidth}{!}{\includegraphics[66,50][424,310]{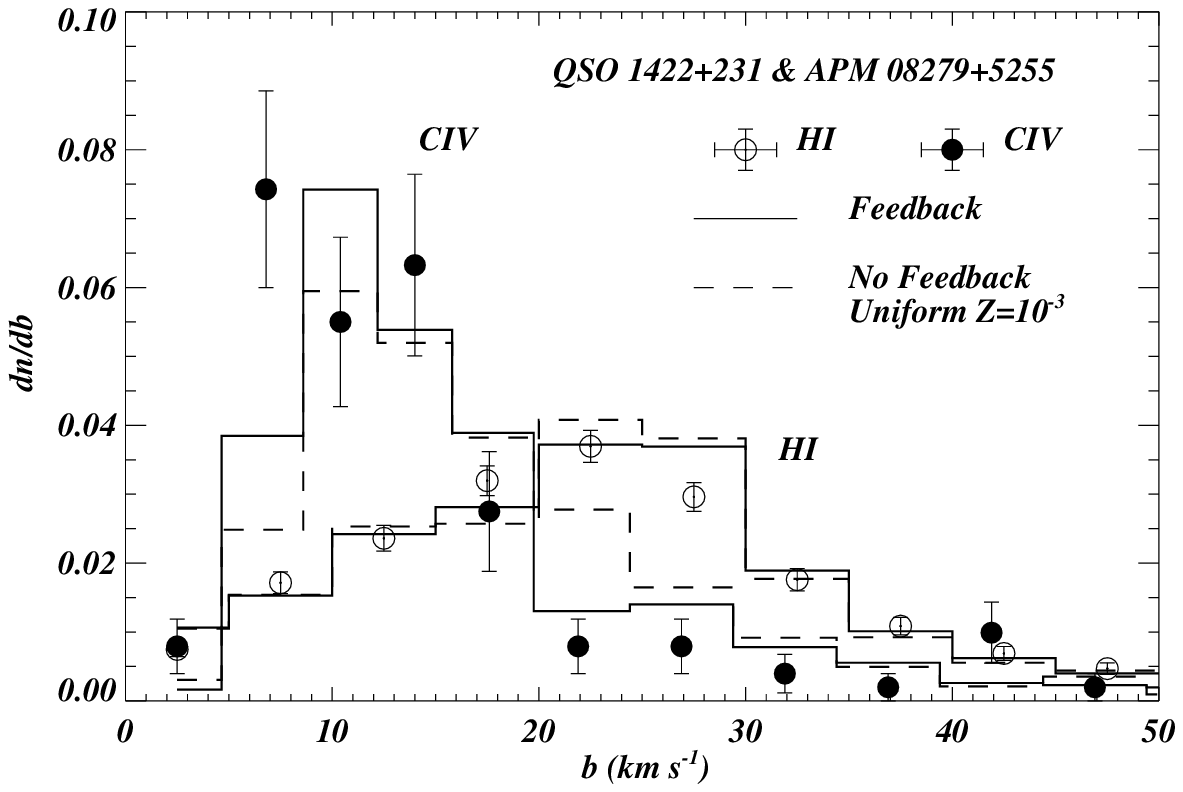}}}
\figcaption{
Line-widths distribution from the combined line-lists
of quasars Q1422+231 (Ellison et al.\ 2000; $z_{\rm em}=3.6$) and APM
08279+5255 ($z_{\rm em}=3.91$) for \Cfour{} (filled circles) and \h{}
(open circles). Full lines are the corresponding results for the
feedback simulation. Dashed lines are the \h{} and \Cfour\ line-width
distributions for the simulation without feedback (and uniform
enrichment). (Only lines for which $12<\log_{10}N_\Cfour{}< 13.2$, and
errors in $N_\Cfour$ and $b_\Cfour$ less than 0.3 dex are included.)
The feedback simulation matches both distributions well, and feedback
has little influence on the \h{} distribution.}
\end{inlinefigure}

Figure~1 illustrates the effect of feedback from star formation on the
density and temperature of the IGM. In the simulation without feedback,
the density structure displays the usual filamentary pattern of mildly
overdense regions, mostly at temperatures $\sim 10^4$~K, although
accretion shocks have also generated some significantly hotter gas.  In
the simulation with feedback, galaxies are surrounded by hot, windblown
bubbles; in particular, the most massive object in the slice is
embedded in a halo of $\approx 1~$\Mpc\ co-moving size, of temperature
up to $10^7$~K.

Except for the very strong absorption lines, the \lya{} spectra are not
significantly affected by these hot bubbles (Fig.~1, bottom panel). In
contrast, feedback has a very strong effect on the stellar masses of
the galaxies (Fig.~2). The hot bubbles disrupt or prevent inflow of
cold gas, which reduces the star-formation rate dramatically.

We have created absorption spectra along random sight lines through the
simulations at $z=3$ and fitted them with Voigt profiles using \vpfit{}
(Webb 1987), mimicking a high signal-to-noise, high resolution spectrum
(see Theuns, Schaye \& Haehnelt 2000 for details of the procedure). We
imposed a power-law ionizing background $J_\nu \propto \nu^\alpha$ with
spectral slope $\alpha= -1.5$ and a break at the \Hep\ Lyman limit (4
Ryd) such that the softness parameter (i.e., the ratio of the \H\ and
\Hep\ ionization rates) $\Gamma_\h/\Gamma_\hep=10^3$. Such a soft
spectrum may be appropriate for $z\ga 3$, when \Hep\ reionization was
incomplete (e.g., Theuns et al.\ 2002 and references therein).  The
amplitude of the ionizing flux is scaled to give the simulations the
observed mean \lya\ absorption. The required flux differs by around 10
per cent between the two simulations, with the simulation with feedback
requiring more ionizing photons because the density in voids is
enhanced by the winds. (Note that this ionizing background differs from
the one imposed during the simulation, see Theuns et al.\ 1998 for tests
of such rescaling.)

The \lya{}-column density distribution functions (CDDFs; the number of
absorption lines per unit column density, per unit redshift) for the
two simulations are compared in Fig.~3a: they are nearly
indistinguishable and fit the observations well\footnote{The
discrepancy between the simulations and the observations for $N_\h >
10^{15.5}~\cm^{-2}$ is caused mainly by the missing large-scale power
due to the finite size of the simulation box.}. The same is true for
the distribution of absorption line widths ($b$-parameters;
Fig.~4). This is the main result of this {\em Letter}: even though the
galaxies in the feedback simulation drive strong winds, there is no
discernible effect on the \lya-forest. Closer examination of the
properties of the hot bubbles reveals why this is so: the winds expand
preferentially into the lower density regions (Fig.~1), and so keep the
filaments that produce the hydrogen lines intact. Only very strong
lines differ noticeably between the two simulations, but this does not
affect the statistics significantly.

Another way to demonstrate that our feedback implementation {\em does}
influence the IGM is by examining the \Cfour{} forest. The hot gas
surrounding the galaxies has been enriched with metals and produces
\Cfour{} absorption. We used \cloudy{} (version 94; Ferland 2000) to
solve the ionization balance equations, computed mock \Cfour{} spectra
(see Aguirre, Schaye \& Theuns 2002 for details) and analyzed them with
\vpfit{}. Fig.~3b shows that the feedback simulation reproduces the
observed \Cfour{} CDDF very well. In contrast, the low and high end of
the CDDF cannot be matched simultaneously by imposing a uniform
metallicity (hashed region).

The metal distribution is very inhomogeneous in the simulation, since
metals are deposited by the winds predominantly in the surroundings of
galaxies. This causes a scatter in the \Cfour{} column density for a
given \lya{} column density. We have proceeded as follows to quantify
this effect. Find the strongest \lya-line in the line-list, and sum up
the column-densities of all \h\ lines, as determined by \vpfit{}, of
which the center falls within $\pm\Delta$\kms. This is the \h\ column
density of that \lq system\rq{}. Use the same algorithm to compute the
\Cfour{} column-density at the corresponding wavelength. Remove all
lines that have contributed to the column-density from the line list,
and start again.  Fig.3c plots \Cfour{} versus \h{} column density for
$\Delta=150$. The feedback simulation reproduces the median and scatter
in the observations relatively well, whereas the uniform metallicity
case underpredicts the scatter (see also Dav\'e et al 1998).

However, some caution is appropriate. Given that in the observations
$N_\cfour/N_\h \approx 10^{-2.6}$, with no evidence for a trend with
$N_\h$ (e.g., Ellison et al.\ 2000), the lack of strong \Cfour\ lines
for the uniform metallicity case may be caused in part by the fact that
the simulation underpredicts the number of strong \H\ lines. Using a
harder UV background, would increase the metallicity required to match
the observations. On the other hand, including metal cooling would
increase the \Cfour\ fraction (and thus strengthen the \Cfour\ lines)
considerably for the feedback simulation. With our choice of
parameters, the feedback simulation reproduces the \Cfour{} data well.
However, other choices of parameters might do equally well.

In the feedback simulation, most of the intergalactic metals reside in
hot ($T \sim 10^{4.5} - 10^{6.5}~\K$), relatively metal-rich ($Z \sim
0.1 - 1.0~Z_\odot$) gas, and collisional ionization has a significant
effect on the strength of metal lines such as \Cfour, \Nfive, and
\Osix. Consequently, the temperature of the gas that dominates the
absorption in one transition, e.g.\ \Cfour{}, can be different from
that of another transition, e.g.\ \h{} or \Osix{}. In such a scenario
one would expect the metal line widths (or $b$-parameters) to be higher
than in the simulation without feedback (and uniform enrichment). This
is in fact not obvious from Fig.~4, because the lines also have a
significant nonthermal component.  At present, both distributions seem
to be in reasonable agreement with the data (filled circles with error
bars). Again, some caution is appropriate: metal cooling (which was not
included) could reduce both the line widths and the importance of
collisional ionization in the feedback simulation. A more detailed
investigation of the metal lines is beyond the scope of this
letter. Here we merely conclude that the feedback simulation predicts a
(highly inhomogeneous) metal distribution that appears to be in
reasonable agreement with the current data, but differs significantly
from the uniform metallicity models considered previously (e.g. Rauch
et al.\ 1997).

{\em In summary}, galactic winds that are strong enough to strongly
suppress galaxy formation and to pollute the IGM with enough metals to
reproduce the data, do not have a significant effect on the
\lya{}-forest. The reason is that the winds fill only a small fraction
of the volume and tend to expand into the voids, leaving the filaments
that produce the hydrogen lines intact.

{\em Acknowledgments} TT thanks PPARC for the award of an Advanced
Fellowship, and M Pettini and M Rauch for comments on the ms. JS
acknowledges support from the W.M.~Keck Foundation and MV from an EARA
Marie Curie Fellowship under contract HPMT-CT 2000-00132. This work was
supported by the European Community Research and Training Network \lq
The Physics of the Intergalactic Medium\rq{} and was conducted in
cooperation with Silicon Graphics/Cray Research utilizing the Origin
2000 super computer at the Department of Applied Mathematics and
Theoretical Physics in Cambridge.

{}

\end{document}